\documentclass[manuscript]{aastex}
\newcommand{\lrab}{{\bf r}_i-{\bf r}_j}
\newcommand{\rab}{{\bf r}_{ij}}
\newcommand{\sab}{r_{ij}}
\newcommand{\rabdot}{\hbox{$\dot {\bf r}_{ij}$}}

\newcommand{\ths}{\thinspace}
\newcommand{\w}{\ \ \ \ \ }
\newcommand{\tle}{{\raise 0.3ex\hbox{$\sc {\ths \le \ths }$}}}
\newcommand{\tge}{{\raise 0.3ex\hbox{$\sc {\ths \ge \ths }$}}}
\newcommand{\tl}{{\raise 0.3ex\hbox{$\sc {\ths < \ths }$}}}
\newcommand{\tg}{{\raise 0.3ex\hbox{$\sc {\ths > \ths }$}}}
\newcommand{\ts}{{\raise 0.3ex\hbox{$\sc {\ths \sim \ths }$}}}
\newcommand{\Msun}{\hbox{$\ths M_{\odot}$}}

\shorttitle{Perturbing forces in planetary systems}
\shortauthors{Kiseleva-Eggleton \& Bois}

\begin{document}
\title{Effects of perturbing forces on the orbital stability of 
planetary systems}
\author{L.Kiseleva-Eggleton\altaffilmark{1}} 
\affil{IGPP, Lawrence Livermore National Laboratory, L-413, 7000 East Ave, Livermore, CA 94550}
\email{lkisseleva@igpp.ucllnl.org}
\and
\author{ E.Bois}
\affil{Observatoire de Bordeaux, 2 rue d'Observatoire, B.P.89, F-33270,
Floirac, France}
\email{bois@observ.u-bordeaux.fr}

\altaffiltext{1}{Also Dept. of Physics, University of California, Davis}

\begin{abstract}
We consider dynamical  effects of additional perturbative forces due to 
the non-point mass nature of stars and planets: effects such as quadrupolar 
distortion 
and tidal friction in the systems of exo-planets. It is shown that these 
forces should not be  neglected while modelling the dynamics of 
planetary systems, especially taking into account the undefined real masses 
of the planets  due to unknown orbital inclinations and the unsatisfactory 
application of Keplerian fits to the radial velocity data in multiple 
planetary systems.
\end{abstract} 


\section{Introduction}

About 50 extrasolar planets have been discovered so far, among them: at least 
three confirmed planetary systems around the stars $\upsilon$ And with three
Jupiter mass planets (e.g. Laughlin \& Adams 1999, 
Rivera \& Lissauer 2000); HD 83443  with two Saturn type 
companions (Mayor et al. 2000); and GJ 876 orbited by two resonant planets 
(Marcy et al. 2001). In addition there are two substellar-mass companions 
orbiting HD 168443 (Marcy et al. 2001, Udry et al. 2001), a system 
of Earth-mass planets around 
pulsar PSR 1257+12 (e.g. Konacki et al. 1998), as well
 as yet-to-be-confirmed systems such as a second planet around 55 Cnc 
(Jayawardhana et al. 2000), planets orbiting Lalande 21185 (Walker 1996), 
and three Earth-type planets around pulsar PSR 1828-11. No 
doubt the number of detected planetary systems will continue to increase
rapidly due to the improvement of observational detection techniques and to new 
space missions (e.g. Kepler, COROT, FAME, SIM etc.) scheduled for the next few 
years.  

The majority of members of new planetary systems have eccentric orbits. 
This may increase the dynamical interaction between  components and make 
systems chaotic and potentially hierarchically unstable, ending up with
an ejection of one or more (in systems with 
more than 2 planets component)  to infinity. Other main factors which add 
to uncertainty of the 
dynamical state of the systems are 
{\bf (a)} the unknown value of orbital inclination 
$i$, which allows us to know only the lower limit of planetary masses $m_o$ from
the function $m_o=m_p\sin i$, and leaves the real mass $m_p$ as almost a 
free parameter (for attempts of $i$ estimation see Gatewood et al. 2001, 
Pourbaix 2001), and 
{\bf (b)} the unknown relative inclination between 
planetary orbital planes. Also, usually the
orbits in planetary systems are  not Keplerian because of mutual 
interaction between planets. Therefore standard Keplerian fits to velocity
observations are  strictly speaking not valid. This point was addressed by 
Laughlin \& Chambers (2001) who suggested  a new four-stage procedure 
for the determination of dynamical 
parameters of multiple planetary systems. This procedure includes 
multiple-Keplerian fits using a semi-analytic scheme followed by a final 
self-consistent polish with N-body interations. This method should be a 
substantial
improvement to existing fitting  techniques. However, in cases when 
close approaches of two or more components are possible, one should 
take into account the non-point-mass nature of the bodies and, as a consequence, 
the following perturbations to Newtonian gravity:
({\bf a}) the quadrupolar distortion (QD) of the bodies due to their mutual 
gravity;
({\bf b}) the further quadrupolar distortion due intrinsic spin of the 
components;
({\bf c}) tidal friction (TF);
({\bf d}) General Relativity.
In this paper we present a few examples showing the influence of  two of 
these perturbations - ({\bf a}) and ({\bf c})- on dynamical stability of systems 
with two planets.

\section{The model}

In order to estimate the dynamical effects of quadrupolar distortion of 
interacting bodies due to their mutual distortion (QD) and to tidal friction
(TF), based here on the near-equlibrium approximation,
on the planetary systems treated as N-body systems with extra forces 
due to QD and TF in addition to the Newtonian gravity, 
we applied the following formulation for the force ${\bf F}_{ij}$ 
of one body on the other, developed by Kiseleva, Eggleton \& Mikkola (1998)
 - hereinafter KEM98:  

\begin{eqnarray}
{\bf 
F}_{ij}=-\left[{Gm_im_j\over\sab^3}+{6G(m_j^2A_i+m_i^2A_j)\over\sab^8}\right]\rab \nonumber \\
+\left[{27\over 2}{\sigma_i m_j^2A_i^2+\sigma_jm_i^2A_j^2\over\sab^{10}}
\ths\rab.\rabdot \right]\rab\w,
\end{eqnarray}
where
\begin{eqnarray}
A_k= {R_k^5Q_k\over 1-Q_k},\w \sigma_k={\alpha_k\over Q_k^2m_kR_k^2}
\sqrt{{Gm_k\over R_k^3}}\w, \nonumber \\
\rab\equiv\lrab.
\end{eqnarray}
Here $R_k$ are the stellar or planetary radii, and $\alpha_k$ are the 
dimensionless
dissipation rates for the two bodies. $Q$ is a version of the apsidal motion 
constant, a dimensionless measure of the distortability of the body 
(star or planet).
KEM found that for a polytrope of index $n$ in the range $0\tle n\tle 4.95$, 
$Q$ can be approximated  by the interpolation formula
\begin{equation}
Q\ \approx\ {3\over 5}\ths\left(1-{n\over 5}\right)^{2.215}\ths 
e^{0.0245n-0.096n^2-0.0084n^3}
\end{equation}

We applied this model to the $\upsilon$ And planetary system. Recently Jiang 
\& Ip (2001) confirmed once more that the innermost planet does not affect 
very much the dynamics of the middle and outer planets, and so we ignored it in 
our 
simulations. For  calculation we used  the regularized 
CHAIN method (Mikkola \& Aarseth 1993) with perturbations. The actual 
numerical integration of the equations of motion
are carried out by a Bulirsh-Stoer integrator with a timestep accuracy of
$10^{-14}$. 

In our simulations we always used  initial orbital parameters 
for orbital periods and eccentricities 
from the Lick Data (Butler et al. 1999, Rivera \& Lissauer 2000): 
$P_c=242$ days, $e_c=0.23$ for the middle planet {\bf C} and
$P_d=1269$ days, $e_d=0.36$ for the outermost planet {\bf D}.
 We always started simulations  with all three components positioned
 in the same plane at apoastrons of their orbits and with 
the two orbits out of phase with each other by $90\deg$ (so we did not 
use the observed values of $\omega$). 
Orbital parameters for the  $\upsilon$ And planetary system 
are not very precisely defined and may vary rather
significantly (particularly the eccentricities) during the lifetime of
the system (see below). Note, however, that differences between values
of orbital parameters given by different authors may not be explained
by real changes during a short observational time of a few years. 
Any real and significant changes require a time scale of hundreds years.  

\section{Results}

In our first set of simulations we use the nominal values of the
planetary masses 
from the Lick Data: for $\sin i = 1$ $m_c=1.98 M_J$, $m_d=4.11 M_J$.
The adopted mass of the star was $M=1.3 \Msun$. The relative inclination 
between the two orbital planes was taken to be $\phi=1\deg$ (models with higher 
relative 
inclinations were also tested but we will discuss those results 
in our next paper). 
For this set of 
parameters the system appear to be hierarchically stable over at 
least $10^7$ years, despite  both orbital eccentricities fluctuating 
almost quasi-periodically within significant ranges: $e_c^{max}-e_c^{min}
\approx 0.5$, $e_d^{max}-e_d^{min}\approx 0.15$. These results are in good 
qualitive agreement with other simulations (Laughlin \& Adams 1999,
Rivera \& Lissauer 2000). Neither TF or QD or their combination 
changes notably (except very minor details) the orbital evolution of
this system. 

However, because of the unknown values of the inclination 
of the orbital planes of the planets $i_c$ and $i_d$, which can differ very much 
from each other, even the mass hierarchy of the $\upsilon$ And planetary system
is questionable. If one accepts the values of $i_c=173.7\deg\pm 3.8\deg$, 
$i_d=28.7\deg \pm 
16.8\deg$ proposed by Pourbaix (2001) with a big uncertainty, this will lead to a 
very different system with component {\bf C} being no longer a planet 
($m_c\approx 
18 M_J$), and $m_d\approx 8.5 M_J$ (so the mass hierarchy is now changed). 
In agreement with earlier stability analyses (Stepinski et al. 2000,
Rivera \& Lissauer 2000) the new system is highly unstable and 
disintegrates ejecting the component {\bf D} within the first 1000 years. 
The addition of QD increases significantly the lifetime of 
the system, although it does not change the chaotic and unstable character of
its dynamics.

\begin{figure}
\caption{Eccentricity of the outer planet in the $\upsilon$ And planetary system 
with masses of planets {\bf C} and {\bf D} exchanged. The system is unstable
without QD and TF (top panel) and with relatively weak QD and TF (second and 
fourth panels), although QD increases the life time of the system. QD with 
coefficients $Q_*=0.08$ and $Q_c=Q_d=0.2$ and TF with $\alpha=10^{-4}$ make 
the system stable over the considered time interval ($10^6$ years), with 
quasi-periodic
fluctuations of $e_d$ (third and bottom panels)} 
\end{figure}

In order to study more systematically the effects of QD and TF 
(in this work we study these effect separately) on dynamical 
stability of planetary systems unstable in point mass approximation, we 
exchanged the masses of planets {\bf C} and {\bf D} in the $\upsilon$ And 
system, 
leaving their intial eccentricities and orbital periods unchanged. Without 
QD and TF this system is unstable ejecting the less massive ($m_d=1.98 M_J$)
component {\bf D} within $\sim 3\times 10^5$ years (top panel of Fig. 1).
QD and TF may significantly change the dynamical evolution of the system. The 
scale of these changes depends very much on the chosen values of coefficients 
$Q_i$ and $\alpha_i$ from Eq 2. The value of $Q_i$ is defined by the 
polytropic
index $n_i$ of the body (Eq 3). For the presumably radiative star of 
$1.3 \Msun$ we assume $n\sim 3$ with corresponding $Q_*\approx 0.02$, and for 
planets we took  $Q\approx 0.2$ ($n\sim 1.5$). The dynamics of the system 
during the first $\sim 1.5\times 10^5$ years is significantly less chaotic than
without QD and the planet D is not ejected to infinity until  $t\sim 5\times 
10^5$ years. But the situation changes really dramatically when we increase
the value of $Q_*$ to 0.08 (so assuming a more convective interior of the star). 
The third panel of Fig. 1 show that in this case the system 
displays
hierarchically stable dynamical behaviour over at least $10^6$ years, 
with quasi-periodic fluctuations of both orbital eccentricities similar to 
ones of original $\upsilon$ And system. 

The dynamical effect of tidal friction  depends on its coefficients 
$\alpha_i$.
Kiseleva et al. (1998) found that for stars similar to the ones in the $\lambda$ Tau 
triple system (Fekel \& Tomkin 1982) the most likely $\alpha\sim 10^{-5}$.
However, for planets $\alpha$ can be significantly larger. We tested our 
model  with $\alpha=10^{-5}$ and $\alpha=10^{-4}$ for all 3 bodies. 
The results shown on the two 
lower panels of Fig. 1 are totally different. $\alpha=10^{-5}$ does not seem 
to improve the stability of the system.
However, for $\alpha=10^{-4}$ the bottom panel presents once more a 
 hierarchically stable system   
with quasi-periodic behaviour of its orbital parameters such as $e_d$. 
Dynamical evolution of models with strong QD ($Q_*=0.08$) 
and with strong TF  ($\alpha=10^{-4}$) over $10^6$ years 
look in this case remarkably similar, despite different dynamical properties of these 
perturbations: TF is a dissipative force with respect to the total
orbital energy and QD is conservative. However, such a similarity does not 
appear in other cases (see below), and we suspect that 
over longer time the evolutionary patterns with QD and with TF will diverge.

We also studed models of  $\upsilon$ And with $\sin i = 0.33$ for both 
 external planets {\bf C} and {\bf D}, so the mass hierarchy of the nominal 
system is preserved and $M_c = 6.53 M_J$, $M_d = 13.56 M_J$. The results are 
shown on Fig.2. TF and especially QD significantly increase the lifetime (upto correspondingly $\sim 5\times 10^4$ and $\sim 1\times 10^5$ yrs) of 
this unstable system.

\begin{figure}
\caption{Eccentricities of planets C (left) and D (right) in $\upsilon$ And with 
$\sin i_c = \sin i_d = 0.33$.
Without TF and QD the system is destroyed within the first 20000 years 
(top panel). TF (middle panel) and especilly QD (bottom panel) significantly 
increase the lifetime of the system. The system with QD actually disintegrates after $\sim 10^5$ yrs.}
\end{figure}
   
\section{Conclusions and prospects}

Our strongest conclusion is that the effects of perturbative forces such as 
quadrupolar distortion and tidal friction should not be neglected when 
investigating numerically  the dynamical properties of extra-Solar planetary
systems, especially when their hierarchical stability is questionable and even
weak additional forces may change the qualitive character of their dynamics. In
our examples QD always had a stabilizing effect, but we can not claim that it 
always works this way and more studies are needed.

 We did not consider here the contribution of General Relativity and intrinsic 
rotation of the bodies which under some circumstances may be important. In 
any case, because of problems with the reliable determination of orbital 
parameters which cannot be consider as Keplerian in N-body systems,
it would be very useful to apply a good dynamical chaos indicator (Lyapunov-type exponent, for example), which can distinguish a long-term dynamical 
instability from  relatively short-term integrations,  given a good-size sample
of possible initial parameters. One such indicator was suggested recently by
Cincotta \& Sim\'o (2000). Our first attempts to apply it to extra-Solar 
planetary systems  are very encouraging and we are going to discuss 
the results in our next paper (Go\'zdziewski, Bois, Maciejewski 
\& Kiseleva-Eggleton, in preparation). 

\acknowledgements
The authors are grateful to the John Templeton Fondation for the grant 
which supported the publication of this paper in \apj.
 LK-E thanks the Bordeaux Observatory for hospitality.


\begin{references}
\reference{}Butler, R.P., Marcy, G.W., Fischer, D. A., Brown, T. M.,
 Contos, A. R., Korzennik, S. G., Nisenson, P. \& Noyes, R. W.,
1999, \apj, 526, 916
\reference{}Cincotta, P. M. \& Sim\'o, C., 2000, A\&ASS, 147, 205
\reference{}Gatewood, G., Han, I. \& Black, D., 2001, \apj, 548, L61
\reference{}Fekel, F. C. \& Tomkin, J., 1982, \apj, 263, 289
\reference{}Jayawardhana R., Holland W., Greaves J., Dent W., Marcy G., 
Hartmann L. \& Fasio G., 2000, \apj., 536, 425   
\reference{}Jiang, I.-G. \& Ip, W.-H., 2001, \apj, submitted
\reference{}Kiseleva, L. G., Eggleton, P. P. \& Mikkola, S., 1998, \mnras,
 300, 292
\reference{}Konacki M., Maciejewcki A. \& Wolszczan A. 1998, \apj. 513, 471
\reference{}Laughlin, G. \& Adams, F. C., 1999, \apj, 526, 881
\reference{}Laughlin, G. \& Chambers, J. E., 2001, \apj, in press
\reference{}Marcy, G.W., Butler, R.P., Fisher, D., Vogt, S., Lissauer, J. J. \& Rivera, E.,
2001, \apj, submitted
\reference{}Marcy, G.W., Butler, R.P., Vogt, S., Liu, M. C., Laughlin, G., Apps, K., 
Graham, J. R., Lloyd, J., Luhman, K. L. \& Jayawardhana, R., 2001, \apj, 
submitted
\reference{}Mayor, M., Naef, D., Pepe, F., Queloz, D., Santos, N. C., Udry, S. \& Burnet,
M., 2000, in Planetary Systems in the Universe, 
 Eds. A. Penny, P. Artymowicz, A.-M. Lagrange and 
S. Russel, ASP Conf. Ser 
\reference{}Mikkola, S. \& Aarseth, S., 1993, Celest.Mech., 57, 439
\reference{}Rivera, E.J. \& Lissauer, J. J., 2000, \apj, 530, 454
\reference{}Pourbaix, D., 2001, A\&A, submitted
\reference{}Stepinski, T. F., Malhotra, R. \& Black, D. C., 2000, \apj, 545,1044
\reference{}Walker, G., 1996, Nature, 382, 23 

\end{references}
\end{document}